\documentstyle[twoside]{article}

\catcode`\@=11
\long\def\@makefntext#1{ 
\protect\noindent \hbox to 3.2pt {\hskip-.9pt
$^{{\eightrm\@thefnmark}}$\hfil}#1\hfill} 

\def\thefootnote{\fnsymbol{footnote}}
 \def\@makefnmark{\hbox to 0pt{$^{\@thefnmark}$\hss}}  

\def\ps@myheadings{\let\@mkboth\@gobbletwo
\def\@oddhead{\hbox{} 
\rightmark\hfil\eightrm\thepage}
\def\@oddfoot{}\def\@evenhead{\eightrm\thepage\hfil 
\leftmark\hbox{}}\def\@evenfoot{}
\def\sectionmark##1{}\def\subsectionmark##1{}}

\oddsidemargin=\evensidemargin
\renewcommand{\thefootnote}{\fnsymbol{footnote}}

\newcounter{sectionc}\newcounter{subsectionc}\newcounter{subsubsectionc}
\renewcommand{\section}[1] {\vspace{12pt}\addtocounter{sectionc}{1}
\setcounter{subsectionc}{0}\setcounter{subsubsectionc}{0}\noindent
	{\tenbf\thesectionc. #1}\par\vspace{5pt}}
\renewcommand{\subsection}[1] {\vspace{12pt}\addtocounter{subsectionc}{1}
	\setcounter{subsubsectionc}{0}\noindent
	{\bf\thesectionc.\thesubsectionc. {\kern1pt \bfit #1}}\par\vspace{5pt}}
\renewcommand{\subsubsection}[1] {\vspace{12pt}\addtocounter{subsubsectionc}{1}
	\noindent{\tenrm\thesectionc.\thesubsectionc.\thesubsubsectionc.
	{\kern1pt \tenit #1}}\par\vspace{5pt}}
\newcommand{\nonumsection}[1] {\vspace{12pt}\noindent{\tenbf #1}
	\par\vspace{5pt}}

\newcounter{appendixc}
\newcounter{subappendixc}[appendixc]
\newcounter{subsubappendixc}[subappendixc]
\renewcommand{\thesubappendixc}{\Alph{appendixc}.\arabic{subappendixc}}
\renewcommand{\thesubsubappendixc}
	{\Alph{appendixc}.\arabic{subappendixc}.\arabic{subsubappendixc}}

\renewcommand{\appendix}[1] {\vspace{12pt}
        \refstepcounter{appendixc}
        \setcounter{figure}{0}
        \setcounter{table}{0}
        \setcounter{lemma}{0}
        \setcounter{theorem}{0}
        \setcounter{corollary}{0}
        \setcounter{definition}{0}
        \setcounter{equation}{0}
        \renewcommand{\thefigure}{\Alph{appendixc}.\arabic{figure}}
        \renewcommand{\thetable}{\Alph{appendixc}.\arabic{table}}
        \renewcommand{\theappendixc}{\Alph{appendixc}}
        \renewcommand{\thelemma}{\Alph{appendixc}.\arabic{lemma}}
        \renewcommand{\thetheorem}{\Alph{appendixc}.\arabic{theorem}}
        \renewcommand{\thedefinition}{\Alph{appendixc}.\arabic{definition}}
        \renewcommand{\thecorollary}{\Alph{appendixc}.\arabic{corollary}}
        \renewcommand{\theequation}{\Alph{appendixc}.\arabic{equation}}
        \noindent{\tenbf Appendix \theappendixc #1}\par\vspace{5pt}}
\newcommand{\subappendix}[1] {\vspace{12pt}
        \refstepcounter{subappendixc}
        \noindent{\bf Appendix \thesubappendixc. {\kern1pt \bfit #1}}
	\par\vspace{5pt}}
\newcommand{\subsubappendix}[1] {\vspace{12pt}
        \refstepcounter{subsubappendixc}
        \noindent{\rm Appendix \thesubsubappendixc. {\kern1pt \tenit #1}}
	\par\vspace{5pt}}

\newcommand{\textlineskip}{\baselineskip=13pt}
\newcommand{\smalllineskip}{\baselineskip=10pt}

\def\eightcirc{
\begin{picture}(0,0)
\put(4.4,1.8){\circle{6.5}}
\end{picture}}
\def\eightcopyright{\eightcirc\kern2.7pt\hbox{\eightrm c}}

\def\abstracts#1#2#3{{
	\centering{\begin{minipage}{4.5in}\baselineskip=10pt\eightrm
	\centerline{ABSTRACT}
	\parindent=0pt #1\par
	\parindent=15pt #2\par
	\parindent=15pt #3
	\end{minipage} }\par}}

\renewenvironment{thebibliography}[1]			
	{\ninerm
	 \baselineskip=11pt				
	 \begin{list}{\arabic{enumi}.}
	{\usecounter{enumi}\setlength{\parsep}{0pt}
	 \setlength{\leftmargin 17pt}{\rightmargin 0pt}	
	 \setlength{\itemsep}{0pt} \settowidth		
	{\labelwidth}{#1.}\sloppy}}{\end{list}}

\newcounter{itemlistc}
\newcounter{romanlistc}
\newcounter{alphlistc}
\newcounter{arabiclistc}

\newcommand{\fcaption}[1]{
        \refstepcounter{figure}
        \setbox\@tempboxa = \hbox{\eightrm Fig.~\thefigure. #1}
        \ifdim \wd\@tempboxa > 5in
           {\begin{center}
        \parbox{5in}{\eightrm \smalllineskip Fig.~\thefigure. #1 }
            \end{center}}
        \else
             {\begin{center}
             {\eightrm Fig.~\thefigure. #1}
              \end{center}}
        \fi}

\newcommand{\tcaption}[1]{
        \refstepcounter{table}
        \setbox\@tempboxa = \hbox{\eightrm Table~\thetable. #1}
        \ifdim \wd\@tempboxa > 5in
           {\begin{center}
        \parbox{5in}{\eightrm\smalllineskip Table~\thetable. #1 }
            \end{center}}
        \else
             {\begin{center}
             {\eightrm Table~\thetable. #1}
              \end{center}}
        \fi}

\def\@citex[#1]#2{\if@filesw\immediate\write\@auxout	
	{\string\citation{#2}}\fi			
\def\@citea{}\@cite{\@for\@citeb:=#2\do			
	{\@citea\def\@citea{,}\@ifundefined		
	{b@\@citeb}{{\bf ?}\@warning
	{Citation `\@citeb' on page \thepage \space undefined}}
	{\csname b@\@citeb\endcsname}}}{#1}}

\newif\if@cghi
\def\cite{\@cghitrue\@ifnextchar [{\@tempswatrue
	\@citex}{\@tempswafalse\@citex[]}}
\def\citelow{\@cghifalse\@ifnextchar [{\@tempswatrue
	\@citex}{\@tempswafalse\@citex[]}}
\def\@cite#1#2{{$\null^{#1}$\if@tempswa\typeout
	{IJCGA warning: optional citation argument
	ignored: `#2'} \fi}}

\def\pmb#1{\setbox0=\hbox{#1}
	\kern-.025em\copy0\kern-\wd0
	\kern.05em\copy0\kern-\wd0
	\kern-.025em\raise.0433em\box0}

\def\fnt#1#2{\footnotetext{\kern-.3em
	{$^{\mbox{\scriptsize #1}}$}{#2}}}

\def\fpage#1{\begingroup
\voffset=.3in
\thispagestyle{empty}\begin{table}[b]\centerline{\footnotesize #1}
	\end{table}\endgroup}

\def\runninghead#1#2{\pagestyle{myheadings}
\markboth{{\eightit{\quad #1}}\hfill}{\hfill{\eightit{#2\quad}}}}

\font\tenbf=cmbx10
\font\tenit=cmti10
\font\tenit=cmti10
\font\bfit=cmbxti10 at 10pt
 1
 1
 1

\font\ninerm=cmr9

\font\eightrm=cmr8
\font\eightit=cmti8

\def\qed{\hbox{${\vcenter{\vbox{                          
   \hrule height 0.4pt\hbox{\vrule width 0.4pt height 6pt
   \kern5pt\vrule width 0.4pt}\hrule height 0.4pt}}}$}}

\runninghead{Thouless number and spin diffusion in quantum Heisenberg
ferromagnets
} {Thouless number and spin diffusion in quantum Heisenberg ferromagnets
}

\begin{document}
\normalsize\textlineskip
{\thispagestyle{empty}
\setcounter{page}{1}

\renewcommand{\thefootnote}{\fnsymbol{footnote}} 
\renewcommand{\theequation}{\arabic{sectionc}.\arabic{equation}}


\vspace*{0.88truein}

\fpage{1}
\centerline{\bf
THOULESS NUMBER AND SPIN DIFFUSION}
\vspace*{0.035truein}
\centerline{\bf IN QUANTUM HEISENBERG FERROMAGNETS}
\vspace*{0.037truein}
\centerline{\footnotesize PETER KOPIETZ}
\vspace*{0.015truein}
\centerline{\footnotesize\it Institut f\H{u}r Theoretische Physik der
Universit\H{a}t
G\H{o}ttingen}
\baselineskip=10pt
\centerline{\footnotesize\it
Bunsenstr. 9, D-37073 G\H{o}ttingen, Germany}
\vspace{0.225truein}

\vspace*{0.21truein}
\abstracts{\noindent
Using an analogy between the conductivity tensor of electronic systems
and the spin stiffness tensor of spin systems,
we introduce the concept of the Thouless number $g_0$ and the dimensionless
frequency
dependent conductance  $g( \omega )$ for quantum spin models.
It is shown that
spin diffusion implies the vanishing of the
Drude peak of $g ( \omega )$, and that the spin diffusion coefficient $D_s$
is proportional to $g_0$.
We develop a new method based on the Thouless number to calculate $D_s$,
and present results for $D_s$ in the
nearest-neighbor quantum Heisenberg ferromagnet at infinite
temperatures for arbitrary dimension $d$ and spin $S$.
}{}{}
\vspace*{-3pt}\textlineskip
\section{Introduction}
\label{sec:intro}
\noindent
According to Thouless\cite{Thouless74} the conductivity of
an electronic
system of linear size $L$ is a measure of the rigidity of the wave functions
with
respect to a twist in the boundary conditions.
This is most clearly seen by writing the
real part of the zero-frequency limit of the dimensionless conductance in $d$
dimensions,
 \begin{equation}
 \tilde{g} ( \omega ) = \frac{L^{d-2} \sigma ( \omega )} {e^2 / \hbar }
 \; \; \; ,
 \label{eq:conductanceel}
 \end{equation}
in the form\cite{Abrahams79,Lee87}
 \begin{equation}
 \tilde{g}_{0} \equiv \lim_{\omega \rightarrow 0} \mbox{Re} \tilde{g} (
 \omega )=\frac{\tilde{E}_{c}}{ \tilde{\Delta} }
 \label{eq:Thouless}
 \; \; \; .
 \end{equation}
Here $\tilde{\Delta}$ is the average spacing between energy levels at the
Fermi energy, and $\tilde{E}_{c}$ is the Thouless energy, which is the typical
fluctuation in energy levels caused by replacing periodic with antiperiodic
boundary conditions.
In a system with a finite diffusion coefficient $D$, the Thouless energy
is given by $\tilde{E}_{c} = \hbar D/ L^{2}$, and can be interpreted as
$\hbar$ divided by
the time taken by a particle to diffuse across a box of side $L$.
The dimensionless quantity $\tilde{g}_{0}$ is
called Thouless number\cite{Abrahams79}. The metallic state
is defined by $\tilde{g}_{0} \gg 1$, i.e.
an electronic system is a metal
if an interval of width $\tilde{E}_{c}$ around the Fermi energy contains many
energy levels.

In the present work we shall show that
$\tilde{g} ( \omega )$ has a precise
dimensionless analog $g ( \omega )$ in quantum spin systems, which can be very
useful for
a better understanding of spin diffusion.
For simplicity we shall focus here on the  spin-$S$ quantum Heisenberg
ferromagnet, but it seems that our results can be generalized to  Heisenberg
antiferromagnets,
and models of itinerant magnetism.
The hamiltonian of the nearest neighbor Heisenberg ferromagnet is given by
 \begin{equation}
 {\cal{H}} = - {J} \sum_{{\bf{r}}} \sum_{\mu = 1}^{d} {\bf{S}}_{\bf{r}} \cdot
 {\bf{S}}_{{\bf{r}} + {\bf{a}}_{\mu} }
 \label{eq:QFM}
 \; \; \; ,
 \end{equation}
where the ${\bf{r}}$-sum over the $N$ sites of a $d$-dimensional
lattice, and ${\bf{a}}_{\mu}$, $\mu = 1, \ldots , d$, are
vectors of length $a$  connecting site ${\bf{r}}$ with
its nearest neighbor in direction $\mu$.
We restrict ourselves to a hypercubic lattice, where ${\bf{a}}_{\mu} \cdot
{\bf{a}}_{\nu} =
a^2 \delta_{\mu \nu}$.
$J > 0$ is the exchange coupling,
and the ${\bf{S}}_{\bf{r}}$ are $SU(2)$ spin operators
satisfying ${\bf{S}}_{\bf{r}}^2 = S(S+1)$.

In their elegant quantum fluids approach to frustrated quantum antiferromagnets
Chandra, Coleman, and Larkin\cite{Chandra90} recently introduced the notion
of the dynamic spin stiffness tensor
$K_{\mu  \nu}^{ ij}$, which
is a tensor both in spin space and in real space.
Here and below the indices $\mu , \nu = 1, \ldots , d$  refer to the
$d$ directions in real space, and $i ,j = x,y,z$ refer to the three components
of the spin operators.
A detailed discussion of the physical meaning
of the spin stiffness tensor  has been given in Ref.\cite{Chandra90},
and will not be repeated here.
Roughly, $K_{\mu \nu}^{ij}$
measures the energy change induced by a space- and time dependent
local twist in the direction of the quantization axis of the spins.
This definition is a generalization of
the static spin stiffness $\rho_s^{0}$, which corresponds to a
time-independent
spiral twist of the quantization axis in the
limit that the wavelength of the spiral becomes infinitely
large\cite{Rudnick78}.
The existence of close analogies between
the spin stiffness of classical Heisenberg models and the conductance
of disordered electrons has first been noticed by
Chakravarty\cite{Chakravarty91}.
In Ref.\cite{Kopietz91} we have derived a spectral representation
for $K_{ \mu \nu }^{i j }$ in quantum Heiseberg ferromagnets,
and pointed out a formal similarity with
the Kubo formula for the conductivity of electrons.
Similar to the current response-kernel of an electronic system defined in the
appendix,
the spin stiffness tensor has a diamagnetic- and a paramagnetic part,
 \begin{equation}
 K_{\mu \nu}^{ij} ( {\bf{k}} , E  ) =
D_{\mu \nu}^{ij} + P_{\mu \nu}^{ij} ( {\bf{k}} ,E)
 \label{eq:DPdef}
 \; \; \; ,
 \end{equation}
where $ k^{-1} $ and $ \hbar / E$ are the  wavelength and time scale
characterizing the local twist of the spin directions.
To make the analogy with the electronic problem
manifest, we define a mass $m_s$ by setting
 \begin{equation}
  \frac{\hbar^2}{ m_{s} a^2 } = J
 \label{eq:msdef}
 \; \; \; .
 \end{equation}
The diamagnetic part of the spin stiffness tensor can then be written as
 \begin{equation}
  D_{\mu \nu }^{ ij} =   - \delta_{\mu \nu} \delta^{ij}
  \frac{\hbar^2}{ m_{s} L^2} \sum_{\bf{r}} <
 {\bf{S}}_{\bf{r}} \cdot {\bf{S}}_{ {\bf{r}} + {\bf{a}}_{\mu} }
 - {S}_{\bf{r}}^{i}
 {S}^{i}_{ {\bf{r}} + {\bf{a}}_{\mu} } >
 \label{eq:Rdiadef}
 \; \; \; ,
 \end{equation}
and the paramagnetic part $P_{\mu \nu}^{ij} ( {\bf{k}} , E )$
has the spectral representation
 \begin{eqnarray}
 P_{\mu \nu }^{ij} ( {\bf{k}} , E )
 & = &    \sum_{n,m} p_{n}
 \left[
 \frac{ < n |  {{J}}_{\mu}^{i} ( {\bf{k}} ) | m>
 <m |  {{J}}_{\nu}^{j} ( -{\bf{k}} ) |
n >}
 { E_{m} - E_{n} - E }
 \right.
 \nonumber
 \\
 &  &  + \left.
 \frac{ < n | {{J}}_{\nu}^{j} ( {\bf{k}} ) | m>
 <m |  {{J}}_{\mu}^{i} ( -{\bf{k}} ) |n >}
 { E_{m} - E_{n} + E }
 \right]
 \label{eq:Rparadef}
 \; \; \; .
 \end{eqnarray}
Here $E_{n}$ and $|n>$ are exact eigenvalues and eigenstates of
Eq.\ref{eq:QFM},
$p_n$ are the Boltzmann factors
$p_{n} = e^{-E_{n}/T}/ \mbox{Tr} [e^{-{\cal{H}}/T}]$,
and ${\bf{J}}_{\mu} = [ J_{\mu}^{x} ,
J_{\mu}^{y} , J_{\mu}^{z} ]$ is the spin current operator in direction
${\bf{a}}_{\mu}$,
 \begin{equation}
 {\bf{J}}_{\mu} ( {\bf{k}})  =   \frac{\hbar^2}{m_{s}L }
 \sum_{\bf{r}} e^{i {\bf{k}} \cdot {\bf{r}}}
 \frac{1}{2a} \left[ {\bf{S}}_{\bf{r}} \times {\bf{S}}_{ {\bf{r}} +
{\bf{a}}_{\mu} }
 - {\bf{S}}_{\bf{r}} \times {\bf{S}}_{ {\bf{r}} - {\bf{a}}_{\mu} }
 \right]
 \label{eq:spincur}
 \; \; \; .
 \end{equation}
The temperature $T$ is measured in units of energy.
In $d$ dimensions one can define $3d$  spin currents,
corresponding to the three spin projections
and the $d$ directions in real space.
In a temperature regime where
the system has long-range spin correlations,
the uniform spin stiffness $\rho_{0}^{s}$ is finite\cite{Rudnick78}.
In terms of the spin stiffness tensor defined above, $\rho^{s}_{0}$
is given by
 \begin{equation}
 \rho^{s}_{0} = - L^{2-d} \lim_{{\bf{k}} \rightarrow 0} \left[ \lim_{\omega
\rightarrow 0}
  K^{xx}_{1 1 } ( {\bf{k}} , \hbar \omega + i 0^{+}) \right]
 \label{eq:rhouniform}
 \; \; \; .
 \end{equation}
In case that the system has a spontaneous magnetization, we shall
choose a coordinate system such that the direction defined by the magnetization
is the $z$-direction in spin-space.
{}From Eqs.\ref{eq:Rdiadef} and \ref{eq:Rparadef} it is clear that
the imaginary part of
$K^{xx}_{1 1 } ( {\bf{k}} , \hbar \omega + i 0^{+})$
vanishes at zero frequency, to that  $\rho_{0}^{s}$ is
real.
At $T=0$ Eq.\ref{eq:rhouniform} reduces
to the familiar result $\rho^{s}_{0} = a^{2-d} JS^2$.
For temperatures above the ordering temperature $T_c$, the diamagnetic- and
paramagnetic
contributions to $\rho^{s}_{0}$ precisely cancel\cite{Kopietz91,Chandra90},
so that $\rho^{s}_{0} = 0$.
In fact, $\rho^{s}_{0}$
is the magnetic analog of the
the long-wavelength limit of the London-Kernel
in an electronic system.
Both quantities describe the appearance of
a long-range rigidity in the wave functions.
An obvious question, which apparently has not been discussed in the literature,
is
whether in a localized  spin model one can also
define the analog of the weight of the Drude peak in the expression for the
conductivity of a metal.
Furthermore, what is the analog of the Thouless number in spin systems?
In the present paper we shall answer these questions.

\section{Thouless number and spin conductance}
\setcounter{equation}{0}
\label{sec:thouless}
\noindent
Guided by the rescaled form of the Kubo formula for electrons
given in the appendix, we define for the spin system the functions $g ( \omega
)$,
$P ( \omega )$, $K ( \omega )$,
and $K^{s} ( {\bf{k}} )$ by simply replacing
in Eqs.\ref{eq:condel},\ref{eq:London},
\ref{eq:Pdimdef} and \ref{eq:rhotildedef}
the rescaled current response kernel $\tilde{K}_{11 }$ by the
spin stiffness tensor $K_{11}^{xx}$,
 \begin{eqnarray}
  {g} ( \omega ) &  =   &
  \lim_{ {\bf{k}} \rightarrow 0}
 \frac{ {K}_{11}^{xx} ( {\bf{k}} , \hbar \omega + i 0^+ ) }{ i ( \hbar \omega +
i 0^{+} ) }
 \label{eq:spinconductance}
 \\
 {P} ( \omega ) & = &
 \lim_{ {\bf{k}} \rightarrow 0}
 \frac{  \mbox{Im} {{K}}_{11}^{xx} ( {\bf{k}} , \hbar \omega + i 0^+ ) }{
\hbar \omega  }
 \label{eq:spinPdimdef}
 \\
 {K} ( \omega ) & =  &
 -  \lim_{{\bf{k}} \rightarrow 0}
 \mbox{Re} K_{1 1}^{xx} ( {\bf{k}} , \hbar \omega + i 0^{+} )
 \label{eq:spindrudedef}
 \\
 {K}^{s} ( {\bf{k}} ) & = & -   \lim_{ \omega \rightarrow 0}
 {K}_{11}^{xx} ( {\bf{k}} , \hbar \omega + i 0^+ )
 \label{eq:spinLondon}
 \; \; \; ,
 \end{eqnarray}
We also define spin analogs $K_0$ and $K^{s}_0$ of the weight of the Drude-peak
and the long-wavelength limit of the London-kernel,
 \begin{eqnarray}
  K_0 &  =   &
  \lim_{ {\omega} \rightarrow 0} K ( \omega )
 \label{eq:drudespinweight}
 \\
 {K}^{s}_0  & = &   \lim_{ {\bf{k}} \rightarrow 0}
 {K}^{s} ( {\bf{k}} )
 \label{eq:spinLondonweight}
 \; \; \; .
 \end{eqnarray}
The function $g ( \omega )$
plays the role of the dimensionless conductance
(see Eq.\ref{eq:condel}), and we shall call
$g ( \omega )$ "spin conductance".
$P ( \omega )$ is the paramagnetic contribution to the real part of $g ( \omega
)$
(see Eq.\ref{eq:Pdimdef}), and the zero-frequency limit of $K ( \omega )$
yields the spin-analog of the weight of the Drude peak
(see Eq.\ref{eq:rhotildedef}).
The wave-vector dependent
function $K^{s} ( {\bf{k}} )$ is related to the length-scale dependent
spin stiffness\cite{Kopietz91} $\rho^{s} ( {\bf{k}} ) =  L^{2-d} K^{s} (
{\bf{k}} )$,
and corresponds to the London-kernel of an electronic system (see
Eq.\ref{eq:London}).
A Schwinger-Boson calculation of $\rho^s ( {\bf{k}} )$
for the two-dimensional quantum ferromagnet has been given in
Ref.\cite{Kopietz91}.
Because  the imaginary part of
$K_{11}^{xx} ( {\bf{k}} , i 0^{+} )$  vanishes, it is not necessary to take the
real part on the right-hand side of Eq.\ref{eq:spinLondon}

We now connect the above definitions with the phenomenon of spin diffusion.
It is easy to obtain the following exact spectral
representations,
 \begin{eqnarray}
 P ( \omega ) & = & \pi \left( \frac{1 - e^{- \hbar \omega / T}}{ \hbar \omega
} \right)
 \lim_{ {\bf{k}} \rightarrow 0}
\sum_{n,m} p_{n}  \delta \left( E_{m} - E_{n} - \hbar \omega \right)
 | < n | J_{1}^{x} ( {\bf{k}} ) | m> |^2
 \nonumber
 \\
 & &
 \label{eq:Pspec}
 \\
 K ( \omega ) & = &
 \frac{ \hbar^2}{m_{s} L^{2}} \sum_{\bf{r}} <
 {{S}}_{\bf{r}}^{z}  {{S}}_{ {\bf{r}} + {\bf{a}}_{1} }^{z}
 + {S}_{\bf{r}}^{y}
 {S}^{y}_{ {\bf{r}} + {\bf{a}}_{1} } >
 \nonumber
 \\
 & -  &  \lim_{ {\bf{k}} \rightarrow 0} \sum_{n,m}
 {\cal{P}} \left\{ \frac{p_{n} - p_{m}}{E_{m} - E_{n} - \hbar \omega} \right\}
 | < n | J_{1}^{x} ( {\bf{k}} ) | m> |^2
 \; \; \; ,
 \label{eq:rhospec}
 \end{eqnarray}
where ${\cal{P}}$ denotes the Cauchy principal value.
General hydrodynamic arguments\cite{Forster75} tell us that spin diffusion
can only exist in the paramagnetic regime $T
\geq T_{c}$.
In this case the low-frequency and long-wavelength
behavior of the dynamic structure factor $S ( {\bf{k}} , \omega )$ is of the
form
 \begin{eqnarray}
 S ( {\bf{k}} , \omega ) & \equiv & 2 \pi \hbar \sum_{n,m}
 p_{n} \delta ( E_{m} - E_{n} - \hbar \omega )
 | < n | {{S}}^{x}_{\bf{k}}  | m> |^2
 \nonumber
 \\
 & = & 2 \chi \left[ \frac{   \hbar \omega}{1 - e^{ - \hbar \omega / T}}
\right]
 \frac{ \hbar D_{s} k^2}{ ( \hbar D_{s} k^2 )^2  + (\hbar \omega )^2}
 \label{eq:dynstruc}
 \; \; \; ,
 \end{eqnarray}
where $D_{s}$ is the spin diffusion coefficient, and
$\chi = T^{-1} \sum_{\bf{r}} < S^{x}_{0} S^{x}_{\bf{r}} >$ is the
uniform susceptibility. The Fourier transform of the spin operators
is defined by
${\bf{S}}_{\bf{k}} = N^{-1/2} \sum_{\bf{r}} e^{i {\bf{k}} \cdot
{\bf{r}} } {\bf{S}}_{\bf{r}}$. The Heisenberg
equation of motion for ${\bf{S}}_{\bf{k}}$ yields
 \begin{equation}
 \hbar \frac{ \partial {\bf{S}}_{\bf{k}} }{\partial t} =
 \frac{J}{\sqrt{N}} \sum_{\bf{q}}
 {\bf{S}}_{\bf{q}} \times {\bf{S}}_{ {\bf{k-q}} }
 \sum_{\mu = 1}^{d} \cos \left[ ( {\bf{k-q}} ) \cdot {\bf{a}}_{\mu} \right]
 \label{eq:Heisenbergmotion2}
 \; \; \; ,
 \end{equation}
where the momentum sum is over the first Brillouin zone.
On the other hand, Fourier transformation of the right-hand side of
Eq.\ref{eq:spincur}
yields
 \begin{equation}
 {\bf{J}}_{\mu} ( {\bf{k}} ) =
 i \frac{Ja}{L} \sum_{\bf{q}}
 {\bf{S}}_{\bf{q}} \times {\bf{S}}_{ {\bf{k-q}} }
 \sin \left[ ({\bf{k-q}} )\cdot {\bf{a}}_{\mu} \right]
 \label{eq:Jmotion}
 \; \; \; ,
 \end{equation}
where we have used $ \hbar^2/ ( m_{s} a^{2} ) = J$.
Combining Eqs.\ref{eq:Heisenbergmotion2} and \ref{eq:Jmotion} and taking
matrix elements we obtain
to leading order in $ | {\bf{k}} \cdot {\bf{a}}_{\mu}  | \ll 1$
 \begin{equation}
   (E_{n} - E_{m} ) <n| {\bf{S}}_{ \bf{k} } | m> = \frac{L}{a\sqrt{N}}
  \sum_{\mu = 1 }^{d}
 <n| {\bf{J}}_{\mu} ( {\bf{k}} ) |m> {\bf{k}} \cdot {\bf{a}}_{\mu}
 \; .
 \label{eq:matrix}
 \end{equation}
Inserting this into Eq.\ref{eq:Pspec} we arrive at
 \begin{equation}
 P ( \omega ) = N \frac{(\hbar \omega )^2}{2 \hbar}
 \left[ \frac{ 1 - e^{- \hbar \omega /T}}{ \hbar \omega } \right] \lim_{
{\bf{k}} \rightarrow 0}
 \frac{S ( {\bf{k}} , \omega )}{ (kL)^2}
 \; \; \; .
 \label{eq:gsres1}
 \end{equation}
Finally, using  the assumption of the diffusive form of $S ( {\bf{k}} , \omega
)$
given in Eq.\ref{eq:dynstruc}, we obtain
 \begin{equation}
 \lim_{\omega \rightarrow 0} P ( \omega ) =
 \frac{N}{L^2} \chi \hbar D_s
 \; \; \; ,
 \label{eq:Presult}
 \end{equation}
To complete the analogy with the electronic system,
we should also proof that spin diffusion
implies the vanishing of the weight of the Drude peak
$K_{0}$ defined in Eq.\ref{eq:drudespinweight}.
To show this,  we use the following trick:
We know that for $T > T_{c}$
the uniform spin stiffness $\rho_{0}^{s}$ vanishes
\cite{Chandra90,Kopietz91}, so that
$K_{0}^{s} = - \lim_{{\bf{k}} \rightarrow 0}
\lim_{\omega \rightarrow 0} K_{11}^{xx} ( {\bf{k}} , \omega ) = 0$.
Hence $K_{0} = K_{0} - K_{0}^{s}$ for $T > T_c$.
But the diamagnetic part $D_{\mu \nu}^{ij}$
of the spin stiffness tensor is independent of momenta and frequency,
and does not appear in the difference $K_{0} - K_{0}^{s}$, so that
 \begin{equation}
 K_{0}  - K_0^{s}  =   \left[
 \lim_{ {\bf{k}} \rightarrow 0} \lim_{\omega \rightarrow 0} -
 \lim_{ \omega \rightarrow 0} \lim_{{\bf{k}} \rightarrow 0}
  \right]
 \sum_{n,m}
 {\cal{P}} \left\{ \frac{p_{n} - p_{m}}{E_{m} - E_{n} - \hbar \omega} \right\}
 | < n | J_{1}^{x} ( {\bf{k}} ) | m> |^2
 \label{eq:KKdif}
 \; \; \; .
 \end{equation}
The contribution from all non-degenerate states $E_{n} \neq E_{m}$
cancels on the right-hand side of this identity, because in this case the
zero-frequency limit is
harmless and we may interchange the order of the limits. Degenerate states
contribute
only if the limit $\omega \rightarrow 0$ is taken before the limit ${\bf{k}}
\rightarrow 0$,
because for finite $\omega$ we have $\lim_{E_m \rightarrow E_{n}} (p_{n} -
p_{n})/
( E_{m} - E_{n} - \hbar \omega ) = 0$. Hence
 \begin{equation}
 K_{0}  - K_0^{s} =
 \frac{1}{T} \sum_{ \stackrel{n,m}{E_{n} = E_{m}}} p_{n}
 | < n | J_{1}^{x} ( {\bf{k}} ) | m> |^2
 \label{eq:KKdif2}
 \; \; \; .
 \end{equation}
Using the identity
 \begin{equation}
 \sum_{ \stackrel{n,m}{E_{n} = E_{m}}} =  \lim_{\omega \rightarrow 0}
 \hbar \int_{- \omega}^{\omega}  d \omega^{\prime}
 \sum_{n,m} \delta ( E_{m} - E_{n} - \hbar \omega^{\prime} )
 \; \; \; ,
 \end{equation}
and comparing the right-hand side of Eq.\ref{eq:KKdif2} with the
spectral representation of
$P ( \omega )$ in Eq.\ref{eq:Pspec}, we conclude that
 \begin{equation}
  K_{0}  =  \frac{ \hbar}{\pi} \lim_{\omega \rightarrow 0}
 \int_{- \omega}^{\omega}  d \omega^{\prime} P ( \omega^\prime )
  + K_0^{s}
 \; \; \; .
 \label{eq:KKP}
 \end{equation}
Using now the fact that
according to
Eq. \ref{eq:Presult} $P ( \omega )$ has a finite
limit as $\omega \rightarrow 0$,
and that $K_0^s = 0$ for $T > T_c$,
we see that the right-hand side of Eq.\ref{eq:KKP} vanishes, so that
$K_{0} = 0$. The vanishing of the Drude peak in the
presence of diffusion is familiar from the electronic
problem, see Eq.\ref{eq:rhotilderes}.
The Thouless number of a spin system with spin diffusion is then
 \begin{equation}
 g_{0} = \lim_{\omega \rightarrow 0} \mbox{Re} g ( \omega ) =
 \frac{E_{c}}{\Delta}
 \; \; \; ,
 \label{eq:gresult}
 \end{equation}
where $\Delta = (N \chi)^{-1}$ plays the role of the level spacing
$ \tilde{\Delta}$ of the electronic system,
and the Thouless energy for the spin system is again defined by $E_{c} = \hbar
D_{s} / L^2$.
Note that the level spacing at the Fermi energy
in an electronic system can be written as $\tilde{\Delta} = (N \kappa)^{-1}$,
where
$\kappa$ is the compressibility.
Comparison with Eq.\ref{eq:Thouless} shows that the
only difference between the Thouless numbers in
electronic- and spin systems is that the compressibility of the
electronic system is replaced by the magnetic susceptibility
of the spin system.

\section{Spin Diffusion}
\setcounter{equation}{0}
\label{sec:apl}

\subsection{General remarks}
\noindent
First of all, it should be emphasized that there exists no proof
of spin diffusion in Heisenberg ferromagnets.
Our calculations in Sec.\ref{sec:thouless} are based on the {\it{assumption}}
that the long-wavelength and low-frequency
behavior of the dynamic structure factor is of the diffusive form given in
Eq.\ref{eq:dynstruc}.
The fact that even for temperatures large compared with $J$
the spin diffusion problem is highly non-trivial
is most clearly seen by writing the spin diffusion
coefficient in the form (see Eqs.\ref{eq:Pspec} and \ref{eq:Presult})
 \begin{equation}
  \hbar D_s a^{-2}  =
 \frac{\pi J}{T \chi}
 \lim_{\Omega \rightarrow 0} \lim_{N \rightarrow \infty}
 \int_{- \infty}^{\infty} d \epsilon
 \frac{\mbox{Tr}_{N} \left\{
 e^{- \beta {\hat{H}} } \hat{I} \delta (\Omega + \epsilon - {\hat{H}} ) \hat{I}
 \delta ( \epsilon - {\hat{H}} ) \right\}}{\mbox{Tr}_{N}
 \left\{ e^{- \beta {\hat{H}} } \right\} }
 \label{eq:spindif}
 \; \; \; ,
 \end{equation}
where $ \beta  =  J / T$,
and $\mbox{Tr}_{N}$ denotes the trace over the Hilbert space of the
$N$-site Heisenberg model, and  the dimensionless operators
$\hat{H} = \hat{H}^{x} +
\hat{H}^{y} +
\hat{H}^{z}$ and
$\hat{I}$ are given by
 \begin{eqnarray}
 \hat{H}^{i} & = &
 - \sum_{{\bf{r}}} \sum_{\mu = 1}^{d} {{S}}_{\bf{r}}^{i}
 {S}_{{\bf{r}} + {\bf{a}}_{\mu} }^{i}
 \; \; \; , \; \; i = x,y,z
 \label{eq:QFM2}
 \\
 \hat{I} & = & \frac{1}{\sqrt{N}} \sum_{\bf{r}} S^{y}_{\bf{r}} \left[
 S^{z}_{ {\bf{r}} + {\bf{a}}_1 } - S^{z}_{ {\bf{r}} - {\bf{a}}_1 } \right]
 \label{eq:jdef}
 \; \; \; .
 \end{eqnarray}
{}From Eq.\ref{eq:spindif} it is obvious
that a finite spin diffusion coefficient can only be obtained
in an infinite system. In any finite system the trace in Eq.\ref{eq:spindif}
will consist of a sum of $\delta$ functions, so that $D_s$
cannot be defined. See Ref.\cite{Thouless74} for a discussion of
this point for the electronic problem.
Moreover,
even after taking the thermodynamic limit $N \rightarrow \infty$,
it is not clear that the right-hand side of Eq.\ref{eq:spindif}
reduces to a finite constant. In principle, there are three
possibilities, familiar from disordered electronic systems:

(1) {\it{Perfect conductor}}. Eq.\ref{eq:spindif} contains two
$\delta$-functions, but only one energy integration. If the current operators
would commute with the Hamiltonian, then the right-hand side of
Eq.\ref{eq:spindif}
would be proportional to $\delta (  \Omega )$, so that
the spin conductance has a Drude-peak, and spin diffusion
does not occur. In this case the system behaves like a perfect conductor.
Although in the Heisenberg model $[ \hat{I} , {\hat{H}} ] \neq 0$,
there exists no proof that the non-commutativity is
sufficient to remove the Drude peak.

(2) {\it{Metal}}.
In this regime the Thouless number $g_0$ and the spin diffusion
coefficient are finite, and are related via Eq.\ref{eq:gresult}.

(3) {\it{Insulator}}.
The third possibility is that in a certain parameter regime
the right-hand side of Eq.\ref{eq:spindif} scales to zero in the
thermodynamic limit. This would correspond to the insulating state
of a disordered electronic system.

\subsection{Spin diffusion at infinite temperature}
\noindent
We now assume that the spin diffusion coefficient
is finite and develop a transparent and direct
method to calculate
$D_{s}$ at infinite temperatures.
The problem of calculating  $D_s$ at $T = \infty$ has been
studied intensely more then 20 years ago\cite{DeGennes58}-\cite{Morita72}.
Most methods are
based on an indirect calculation of $D_s$ via the dynamic structure
factor $S ( {\bf{k}} , \omega )$, assuming that its
long-wavelength and low-frequency behavior is
of the form given in Eq.\ref{eq:dynstruc}.
The spin diffusion coefficient
is obtained indirectly from $S ( {\bf{k}} , \omega )$
by means of the limiting procedure
$\lim_{\omega \rightarrow 0} \lim_{ {\bf{k}} \rightarrow 0}
( \omega^2 / k^2 ) S ( {\bf{k}} , \omega
)$, see Eq.\ref{eq:dynstruc}.

In praxis, it is impossible to calculate $S ( {\bf{k}} , \omega )$
at low-frequencies, or equivalently its real-time
Fourier transform $S ( {\bf{k}} , t )$ at long times $t$.
The moment method first applied
by de Gennes\cite{DeGennes58} is equivalent to
an extrapolation of a short time expansion to long times.
The concept of the Thouless number in spin systems offers a  more
direct way to calculate $D_s$.
Of course, if $S ( {\bf{k}} , \omega )$ could be calculated exactly, then the
result
for $D_s$ would be identical with the result obtained by means an exact
calculation of $g ( \omega )$. However, extrapolations
of high-frequency expansions of $g( \omega )$ and
$S ( {\bf{k}}, \omega )$ will in general not agree,
because only in the limit $\omega \rightarrow 0$
and ${\bf{k}} \rightarrow 0$
there is a direct connection between these two quantities.
We believe that our approach via
the Thouless number $g_0$ is more reliable than extrapolations based on the
dynamic
structure factor, because $g_0$ is directly
proportional  to $D_s$, and no further limiting procedures are required.

We now use the
first two expansion coefficients in the
short time expansion of the right-hand side of Eq.\ref{eq:spindif} to
estimate $D_{s}$ at $T = \infty$.
Introducing  Fourier representations of the $\delta$-functions
and defining $\hat{I} (t) = e^{i \hat{H} t} \hat{I} e^{- i \hat{H} t}$,
we obtain from Eq.\ref{eq:spindif} after straightforward manipulations
 \begin{eqnarray}
  \hbar D_s a^{-2}  & =  &
 \frac{ J}{ T \chi} \int_{0}^{\infty}  C (t)
 \label{eq:Dtdef}
 \\
 C (t) & = & \frac{1}{2}
 < \hat{I}(0)  \left[ \hat{I}(t) + \hat{I} ( -t ) \right] >
 \; \; \; ,
 \label{eq:Cdef}
 \end{eqnarray}
where $< \cdots >$ denotes thermal average with the Hamiltonian $\hat{H}$.
The existence of the integral in Eq.\ref{eq:Cdef} implies
spin diffusion. The convergence of the integral is determined by the
long-time behavior of $C(t)$. Unfortunately, there exists no completely
controlled method to calculate $C(t)$ for large $t$.
We therefore assume that the integral exists, and try to extract
the long-time behavior from the short-time expansion,
 \begin{equation}
C (t) = \sum_{n = 0}^{\infty} \frac{ (-1)^{n} t^{2n} }{(2n) !} C_{2n}
 \label{eq:shortexpansion}
 \; \; \; .
 \end{equation}
Note that only even powers of $t$ appear, because $C(t) = C (-t)$.
The expansion coefficients $C_{2n}$ can be written in terms of multiple
commutators.
The first two coefficients are
 \begin{eqnarray}
 C_{0} & = & < \hat{I}^{2} >
 \label{eq:C0def}
 \\
 C_{2} & = & < \hat{I} \left[ \left[ \hat{I} , \hat{H} \right] , \hat{H}
\right] >
 \label{eq:C2def}
 \; \; \; ,
 \end{eqnarray}
where all operators are at equal times.
At $ T = \infty$ the evaluation of the thermal averages in Eqs.\ref{eq:C0def}
and
\ref{eq:C2def} simplify, because spins at different sites are not correlated.
To calculate $C_{0}$, we need $< (S^{i})^2> = S (S+1)/3$, for
$i = x,y,z$. A short calculation gives
 \begin{equation}
 C_{0} = 2 \left[ \frac{S (S+1)}{3} \right]^2
 \; \; \; .
 \label{eq:C0res}
 \end{equation}
The evaluation of $C_{2}$ is tedious but not difficult.
It involves the expectation values
of up to four spins, which have been tabulated in Ref.\cite{Ambler62}.
The following averages are needed
 \begin{eqnarray}
 < S^{i} S^{j} S^{k} > & = & \epsilon^{ijk}
 \frac{i}{6} {S ( S+1)}
 \label{eq:Sxyz}
 \\
 < (S^{i})^4 > & = & \frac{S(S+1)}{15} \left[ 3 {S(S+1)} -
 1 \right]
 \label{eq:Sxxxx}
 \; \; \; ,
 \end{eqnarray}
 where $< \ldots > = \mbox{Tr} [ \ldots ] / ( 2S+1 )$, and the trace is over
 the $2S+1$ states of the spin-$S$ Hilbert space.
$\epsilon^{ijk}$ is the antisymmetric Levi-Civita tensor.
For $i \neq j$ we have
 \begin{eqnarray}
 < (S^{i})^2 (S^{j})^2> & = &
 \frac{S(S+1)}{15} \left[ { S(S+1) } + \frac{1}{2} \right]
 \; \; \; ,
 \label{eq:Sxxyy}
 \\
 < S^{i} S^{j} S^{i} S^{j} > & = &
 \frac{S(S+1)}{15} \left[ { S(S+1) } - {2} \right]
 \label{eq:Sxyxy}
 \; \; \; .
 \end{eqnarray}
Writing
\begin{equation}
 C_{2}  =  \sum_{ i,j = x,y,z } C_{2}^{ij}
 =
 \sum_{ i,j = x,y,z }
 < \hat{I} \left[ \left[ \hat{I} , \hat{H}^{i} \right] , \hat{H}^{j} \right] >
 \label{eq:C2alphabeta}
 \; \; \; .
\end{equation}
we obtain for a $d$-dimensional hypercubic lattice
 \begin{eqnarray}
 C_{2}^{xx} & = & \left[ \frac{S(S+1)}{3} \right]^2
 \left[ \left( 2 d - \frac{2}{5} \right) \frac{4S(S+1)}{3} - \frac{3}{5}
\right]
 \label{eq:C2xx}
 \\
 C_{2}^{yy} & = & C_{2}^{zz}  =  \left[ \frac{S(S+1)}{3} \right]^2
 \left[ \left(  d - \frac{3}{5} \right) \frac{4S(S+1)}{3} - \frac{2}{5} \right]
 \label{eq:C2yy}
 \\
 C_{2}^{yz} & = & C_{2}^{zy}  =  \left[ \frac{S(S+1)}{3} \right]^2
 \left[ - \frac{4S(S+1)}{15} + \frac{1}{5} \right]
 \; \; \; ,
 \label{eq:C2yz}
 \end{eqnarray}
and $C_{2}^{xy} = C_{2}^{yx} = C_{2}^{xz} = C_{2}^{zx} = 0$.
Hence
 \begin{equation}
 C_{2}= \left[ \frac{S(S+1)}{3} \right]^2
 \left[ \left( 4 d - 2 \right) \frac{4S(S+1)}{3} - 1 \right]
 \label{eq:C2res}
 \; \; \; .
 \end{equation}
We are now ready to extrapolate $C (t )$ to long times.
This extrapolation is of course not unique. A widely used extrapolation scheme,
which we shall follow here,
is to assume that the first two coefficients are consistent with a
Gaussian\cite{Moriya56}. This leads to
 \begin{equation}
 C (t) \approx  C_{0} \exp \left[ - \frac{C_{2} t^2}{2 C_{0}} \right]
 \label{eq:Gausapprox}
 \; \; \; .
 \end{equation}
Using the fact that $T \chi = S (S+1) / 3$ at $T = \infty$, we finally obtain
from
Eqs.\ref{eq:Dtdef}, \ref{eq:C0res}, \ref{eq:C2res} and \ref{eq:Gausapprox}
for the spin diffusion coefficient at $T = \infty$
 \begin{equation}
 \frac{\hbar D_{s}}{a^2 J} = \left[  \frac{ \pi S(S+1)}{3} \right]^{1/2}
 \left[  4d-2 - \frac{3 }{4S(S+1)}   \right]^{-1/2}
 \label{eq:final}
  \; \; \; .
 \end{equation}
This is the main result of this section.
Note that the term proportional to $[S(S+1)]^{-1}$ in the second factor
of Eq.\ref{eq:final} can be viewed as a quantum correction, that becomes
irrelevant in the limit of large $S$.
For $S= 1/2$ Eq.\ref{eq:final} reduces to
 \begin{equation}
 \frac{\hbar D_{s}}{a^2 J} = \frac{1}{2} \left[ \frac{\pi}{ 4d-3} \right]^{1/2}
 = \left\{
 \begin{array}{ll}
 0.886 & \mbox{for $ d = 1$} \\
 0.396 & \mbox{for $ d = 2$} \\
 0.295 & \mbox{for $ d = 3$}
 \end{array}
 \right.
 \label{eq:shalf}
 \end{equation}
The classical limit of the Heisenberg model should be taken by
letting $S \rightarrow \infty$ while keeping $J_{cl} = JS^2$ constant.
In this limit $D_{s}$ vanishes. The leading coefficient for large $S$
can be read off from Eq.\ref{eq:final}.
 \begin{equation}
 \frac{\hbar D_{s}}{a^2 J_{cl}} =
 \frac{1}{ S \sqrt{3}  } \left[ \frac{\pi}{ 4d-2} \right]^{1/2}
 = \frac{1}{S} \times \left\{
 \begin{array}{ll}
 0.724 & \mbox{for $ d = 1$} \\
 0.418 & \mbox{for $ d = 2$} \\
 0.324 & \mbox{for $ d = 3$}
 \end{array}
 \right.
 \label{eq:sinf}
 \end{equation}
Eq.\ref{eq:shalf} agrees with the results listed in the
first column of table V of Ref.\cite{Morita72}.
Hence, at least to second
order in the short time expansion, our method is equivalent
with Morita's memory function formalism\cite{Morita72}.
However, it seems that our Thouless-number approach is physically more
transparent,
because it does not involve the construction of momentum dependent auxiliary
quantities.
The results in $d=3$ are also in excellent agreement with Bennett and
Martin\cite{Bennett65},
who obtained by means
of an indirect extrapolation method based on the
dynamic structure factor ${\hbar D_{s}} / ({a^2 J}) \approx 0.33
\sqrt{S(S+1)}$.
This gives ${\hbar D_{s}} / ({a^2 J}) = 0.29$
for $S=1/2$, and ${\hbar D_{s}} / ({a^2 J_{cl}}) = 0.33/S$
for $S \rightarrow \infty$.

\section{Conclusions and open problems}
\setcounter{equation}{0}
\label{sec:con}
\noindent
In this paper we have used an analogy between the
Kubo formula for disordered electrons
and the dynamic spin stiffness tensor of localized quantum spin models
to define several new quantities that characterize the dynamic behavior
of Heisenberg ferromagnets in the paramagnetic regime.
The key observation is that, after proper rescaling, the dynamic spin stiffness
tensor of
the spin system and the current response kernel of the electronic system
can be written in a formally identical way. The proper definition of the
Thouless number and the dimensionless conductance in the spin system
directly follow from this analogy.
As a first application, we have used the Thouless number to develop
a simple extrapolation scheme for the calculation of the spin diffusion
coefficient
at infinite temperatures.

This work opens a number of new directions for further research:
A generalization of the concepts developed here to antiferromagnets or
Hubbard models seems possible, although some technical difficulties might be
encountered.
The analogy between electrons in the presence of disorder and
spin models in the paramagnetic regime
might serve as a useful guide to understand
spin diffusion in two dimensions.
A Schwinger-Boson calculation of $D_s$ in two-dimensional Heisenberg models
at low temperatures has recently been given by Chubukov\cite{Chubukov91}.
His result for $D_s$ in ferromagnets can be easily
reproduced if we translate the Drude formula
for the dimensionless conductance $\tilde{g}_0$ of an electronic system
into spin language.
By comparing the expressions for the diamagnetic tensors $D_{\mu \nu}^{ij}$ and
$\tilde{D}_{\mu \nu}$ given in Eqs.\ref{eq:Rdiadef} and \ref{eq:Kdiadef},
we conclude that the energy $\hbar^2 N / ( m_e L^2 )$ corresponds to
$\hbar^2 N S^2 / ( m_s L^2)$ in the spin model.
Here we have used the fact that for $T \ll J S^2$ the spin-correlation
length is exponentially large compared with the lattice
spacing\cite{Kopietz89}, so that
the summation in Eq.\ref{eq:Rdiadef} yields $NS^2$\cite{Kopietz91}.
Hence, to obtain the Drude result for $D_s$,
we should replace $ \hbar^2 N / ( m_e L^2 ) \rightarrow
\hbar^2 N S^2 / ( m_s L^2 ) = J S^2$ in Eq.\ref{eq:DrudeThouless}.
This gives for the Thouless number in the Drude approximation
 \begin{equation}
 g_0 = JS^2 \frac{\tau}{\hbar}
 \; \; \; .
 \label{eq:spinconddrude}
 \end{equation}
Here $\tau$ is the characteristic lifetime of single-particle
excitations at wavelengths
large compared with the spin correlation length.
To determine $\tau$, a microscopic calculation is necessary\cite{Chubukov91}.
Combining Eqs.\ref{eq:spinconddrude} and \ref{eq:gresult}, we conclude
that the spin diffusion coefficient in $d=2$ is at low temperatures
given by
\begin{equation}
\hbar D_s a^{-2} = \frac{J S^2 \tau}{\chi \hbar }
\; \; \; ,
\label{eq:spindifres1}
\end{equation}
Eq.\ref{eq:spindifres1} agrees exactly with the result of
Chubukov\cite{Chubukov91}
(see his Eqs.2.30 and 2.32), who performed a diagrammatic
calculation within the Schwinger-Boson formalism.
The factor of $JS^2$ in the enumerator is interpreted by Chubukov as transport
coefficient.
Note, however, that in $d=2$ the Thouless number for
non-interacting electrons in the presence of disorder
scales to zero in the thermodynamic limit\cite{Abrahams79},
so that the system is an insulator.
Assuming that our  analogy can also be applied to this case,
we speculate that for any $T > 0$ the two-dimensional Heisenberg model
corresponds to an insulator, so that the spin diffusion coefficient vanishes.
In any case, we expect that in two dimensions
the interaction between the diffusion modes will be very important,
so that the Drude result given in Eq.\ref{eq:spindifres1} cannot be trusted.
Non-perturbative methods are necessary to
understand spin diffusion in the low temperature regime of two dimensional
Heisenberg models.
%

\nonumsection{References}

\nonumsection{Appendix}
\setcounter{equation}{0}
\renewcommand{\theequation}{A.\arabic{equation}}
\noindent
In this appendix we shall
rewrite the standard Kubo formula\cite{Mahan81} for the
conductivity of an electronic system in a rescaled form.
By comparing the rescaled form of the current-response Kernel
$\tilde{K}_{\mu \nu}$
with the spin stiffness tensor
${K}_{\mu \nu}^{ij}$ defined in Sec.\ref{sec:intro},
the proper definition of the Thouless number
and the dimensionless spin conductance become obvious.
Except for a tilde, we shall use the same symbols as in the spin problem
to emphasize the close similarity between
the rescaled current-response kernel and spin stiffness tensor.

Consider a $d$-dimensional box of volume $L^d$ containing $N$ electrons with
mass $m_{e}$.
Within linear response theory, the change in the current density
$\delta j_{\mu} ( {\bf{r}} , t )$ due to a change in the vector potential
$\delta A_{\nu} ( {\bf{r}}^{\prime}, t^{\prime})$ is given by
 \begin{equation}
 \delta j_{\mu} ( {\bf{r}} , t )
  =   ( {\alpha L^{2-d}}/{\hbar})
 \int_{- \infty}^{\infty} dt^{\prime} \int d {\bf{r}}^{\prime}
 \sum_{\nu}
 \tilde{K}_{\mu  \nu } ( {\bf{r}} , {\bf{r}}^{\prime} , t-t^{\prime} )
 \delta A_{\nu} ( {\bf{r}}^{\prime} , t^{\prime} )
 \label{eq:linearresponse}
 \; \; \; ,
 \end{equation}
where $\alpha = e^2 / (\hbar c) \approx 1/137$ is the fine-structure
constant.
For reasons obvious below, we have not included the factor $\alpha L^{2-d}/
\hbar$
into the definition of $\tilde{K}_{\mu \nu}$.
The Kubo formula for the Fourier transform of the
current response kernel\cite{Mahan81} yields
 $
 \tilde{K}_{\mu \nu}  ( {\bf{k}} , E ) =
 \tilde{D}_{\mu \nu} + \tilde{P}_{\mu \nu} ( {\bf{k}} , E )  $, where
the diamagnetic- and paramagnetic parts are given by
 \begin{eqnarray}
 \tilde{D}_{\mu \nu} & =  & - \delta_{\mu \nu}
  \frac{\hbar^2}{m_{e} L^2} N
 \label{eq:Kdiadef}
 \\
 \tilde{P}_{\mu \nu} ( {\bf{k}} , E )
 & = &    \sum_{n,m} p_{n}
 \left[
 \frac{ < n | \tilde{J}_{\mu} ( {\bf{k}} ) | m><m | \tilde{J}_{\nu} ( -{\bf{k}}
) | n >}
 { E_{m} - E_{n} - E }
 \right.
 \nonumber
 \\
 &  & + \left.
 \frac{ < n | \tilde{J}_{\nu} ( {\bf{k}} ) | m><m | \tilde{J}_{\mu} ( -{\bf{k}}
) | n >}
 { E_{m} - E_{n} + E }
 \right]
 \label{eq:Kparadef}
 \; \; \; ,
 \end{eqnarray}
and the current operators $\tilde{J}_{\mu} ( {\bf{k}} ) $
are defined by
 \begin{equation}
 \tilde{J}_{\mu} ( {\bf{k}} )
 =
 \frac{\hbar^2}{m_{e} L} \int d {\bf{r}} e^{ i {\bf{k}} \cdot  {\bf{r}} }
 \frac{1}{2i} \left[ \hat{\psi}^{\dagger} ( {\bf{r}} )  \frac{ \partial
}{\partial x^{\mu} }
 \hat{\psi} ( {\bf{r}} ) - h.c. \right]
 \label{eq:electoncurdef}
 \; \; \; ,
 \end{equation}
Here $\hat{\psi} ( {\bf{r}} )$ is the usual second quantized field operator for
the electrons.
Note that with this rescaling the $\tilde{J}_{\mu}$ has units of energy, just
like the
spin currents defined in Eq.\ref{eq:spincur}.
The dimensionless conductance $\tilde{g} ( \omega ) = L^{d-2} \sigma ( \omega )
/ (e^2/ \hbar )$,
and the (rescaled) London kernel $\tilde{K}^{s} ( {\bf{k}} )$ are given by
 \begin{eqnarray}
  \tilde{g} ( \omega ) &  =   &
  \lim_{ {\bf{k}} \rightarrow 0}
 \frac{ \tilde{K}_{11} ( {\bf{k}} , \hbar \omega + i 0^+ ) }{ i ( \hbar \omega
+ i 0^{+} ) }
 \label{eq:condel}
 \\
 \tilde{K}^{s} ( {\bf{k}} ) & = & -   \lim_{ \omega \rightarrow 0}
 \tilde{K}_{11} ( {\bf{k}} , \hbar \omega + i 0^+ )
 \label{eq:London}
 \; \; \; .
 \end{eqnarray}
{}From Eq.\ref{eq:condel} it is clear that the
factor $\alpha L^{d-2} / \hbar $ in Eq.\ref{eq:linearresponse}
is very natural.
Note that the "$\omega$-limit" (where the limit $\omega \rightarrow 0$ is taken
after the limit $k \rightarrow 0$)
and the "$k$-limit" (where the limit $k \rightarrow 0$ is taken
after the limit $\omega  \rightarrow 0$)
describe very different physical properties of the system.
The energy $\tilde{K}^{s}_{0} = \lim_{ {\bf{k}} \rightarrow 0} \tilde{K}^{s} (
{\bf{k}} )$ is
proportional to the density of superconducting electrons.
Comparing of Eqs.\ref{eq:rhouniform}
and Eqs.\ref{eq:London}, it is evident that $L^{2-d} \tilde{K}^{s}_{0}$
corresponds to $\rho^s_0$ in the spin system.  Both quantities are only finite
in the presence of off-diagonal long-range order.
{}From Eq.\ref{eq:condel} we obtain
 \begin{equation}
 \mbox{Re} \tilde{g} ( \omega )  =  \pi \tilde{K}_{0} \delta ( \hbar \omega )
 + \tilde{P} ( \omega )
 \; \; \; , \; \; \;
 \mbox{Im} \tilde{g} ( \omega )  =   \frac{ \tilde{K} ( \omega ) }{ \hbar
\omega }
 \; \; \; ,
 \end{equation}
with
 \begin{eqnarray}
 \tilde{P} ( \omega ) & = &
 \lim_{ {\bf{k}} \rightarrow 0}
 \frac{  \mbox{Im} \tilde{{K}}_{11} ( {\bf{k}} , \hbar \omega + i 0^+ ) }{
\hbar \omega  }
 \label{eq:Pdimdef}
 \\
 \tilde{K} ( \omega ) & =  &
 -  \lim_{{\bf{k}} \rightarrow 0}
 \mbox{Re} \tilde{K}_{1 1 } ( {\bf{k}} , \hbar \omega + i 0^{+} )
 \label{eq:rhotildedef}
 \; \; \; ,
 \end{eqnarray}
and $\tilde{K}_{0} = \lim_{\omega \rightarrow 0} \tilde{K} ( \omega )$.
It is instuctive to evaluate the above quantities
in the simplest possible approximation, where
all scattering processes are taken into account
by introducing a phenomenological lifetime $\tau$
in the electronic Greens functions.
At temperatures small compared with the Fermi energy,
this yields the well known Drude results\cite{Ashcroft76}
 \begin{eqnarray}
 \tilde{g} ( \omega ) & = &    \frac{ \hbar N }{m_{e} L^2}
 \frac{  \tau  }{ 1  -  i  \omega \tau }
 \label{eq:gtilderes}
 \\
 \tilde{P} ( \omega ) & = &    \frac{ \hbar N}{m_{e} L^2}
 \frac{  \tau   }{ 1  +  (  \omega \tau )^2}
 \label{eq:Ptilderes}
 \\
 \tilde{K}  ( \omega ) & = &
 \frac{\hbar^2 N}{m_{e} L^2} \left[ 1 - \frac{1}{ 1 + ( \omega \tau )^2}
 \right]
 \label{eq:rhotilderes}
 \; \; \; .
 \end{eqnarray}
The  Thouless number $\tilde{g}_0$ defined in Eq.\ref{eq:Thouless}
is in this approximation given by
 \begin{equation}
 \tilde{g}_0 =  \frac{\hbar N \tau } { m_e  L^2}
 \; \; \; .
 \label{eq:DrudeThouless}
 \end{equation}
Combining Eqs.\ref{eq:Thouless} and \ref{eq:DrudeThouless}, and using
the fact that the level
spacing at the Fermi energy in $d$ dimensions
can be expressed in terms of the Fermi velocity $v_F$ as
$\tilde{\Delta} = m_e v_F^2 / (dN)$,
we obtain the classical result for the diffusion coefficient $D = v_F^2 \tau /
d$.


\begin{thebibliography}{99}
\bibitem{Morita72}
T. Morita, Phys. Rev. {\bf{B 6}}, 3385 (1972).
Note that the definition of $J$ used in this work differs
from the conventional definition by a factor of $2$, so that the numbers
for $\hbar D_s a^{-2} / J$ given by Morita should be divided by $2$ for
a comparison with our result.
\end{thebibliography}
\end{document}